\documentclass[runningheads]{llncs}
\usepackage[T1]{fontenc}
\usepackage{graphicx}
\usepackage{hyperref}
\begin{document}
\title{CXR-FL: Deep Learning-based Chest X-ray Image Analysis Using Federated Learning}
\titlerunning{CXR-FL:  Chest X-ray Image Analysis Using Federated Learning}
%
\author{Filip Ślazyk\inst{1, 2}\orcidID{0000-0002-7270-7836} \and Przemysław Jabłecki\inst{1, 2}\orcidID{0000-0002-7306-1467} \and Aneta Lisowska\inst{1}\orcidID{0000-0002-4489-5956} \and Maciej Malawski\inst{1, 2}\orcidID{0000-0001-6005-0243}  \and Szymon Płotka \inst{1, 3}\orcidID{0000-0001-9411-820X}
}
\authorrunning{F. Ślazyk et al.}
\institute{Sano Centre for Computational Medicine, Krakow, Poland \and
AGH University of Science and Technology, Krakow, Poland \and 
Warsaw University of Technology, Warsaw, Poland}
%
\maketitle              
\begin{abstract}

Federated learning enables building a shared model from multicentre data while storing the training data locally for privacy.
In this paper, we present an evaluation (called CXR-FL) of deep learning-based models for chest X-ray image analysis using the federated learning method. We examine the impact of federated learning parameters on the performance of central models. Additionally, we show that classification models perform worse if trained on a region of interest reduced to segmentation of the lung compared to the full image. However, focusing training of the classification model on the lung area may result in improved pathology interpretability during inference. We also find that federated learning helps maintain model generalizability. The pre-trained weights and code are publicly available at (\url{https://github.com/SanoScience/CXR-FL}).

\end{abstract}

\section{Introduction}

Federated Learning (FL) is an effective privacy-preserving machine learning technique used to train models across multiple decentralized devices. It enables using a large amount of labeled data in a secure and privacy-preserving process \cite{kaissis2020secure} to improve the generalizability of the model \cite{Sheller}. Recent work on the application of federated learning in medical imaging shows promising results in dermoscopic diagnosis \cite{Chen}, volumetric segmentation \cite{Wu} and chest X-ray image analysis \cite{Dong}. In this paper, we evaluate the application of deep learning-based models to medical image analysis using the FL method. To gain insight into the impact of FL-related parameters on the global model, we conduct experiments with a variable number of clients and local training epochs. We explore utilisation of cascading approach, where medical image segmentation is performed prior to classification, for increased pathology classification interpretability. We compare our results with \cite{Teixeira} in terms of explainable AI (XAI) and classification performance. We find faster convergence of the learning process for a greater fraction of selected clients and a greater number of local epochs in the segmentation task. We show that federated learning improves the generalizability of the model and helps avoid overfitting in the classification task. We show that Grad-CAM explanations for classification models trained on segmented images may be more focused on the lung area than those trained on full images.

\begin{figure}[t!]%
    \centering
    \includegraphics[width=0.8\textwidth]{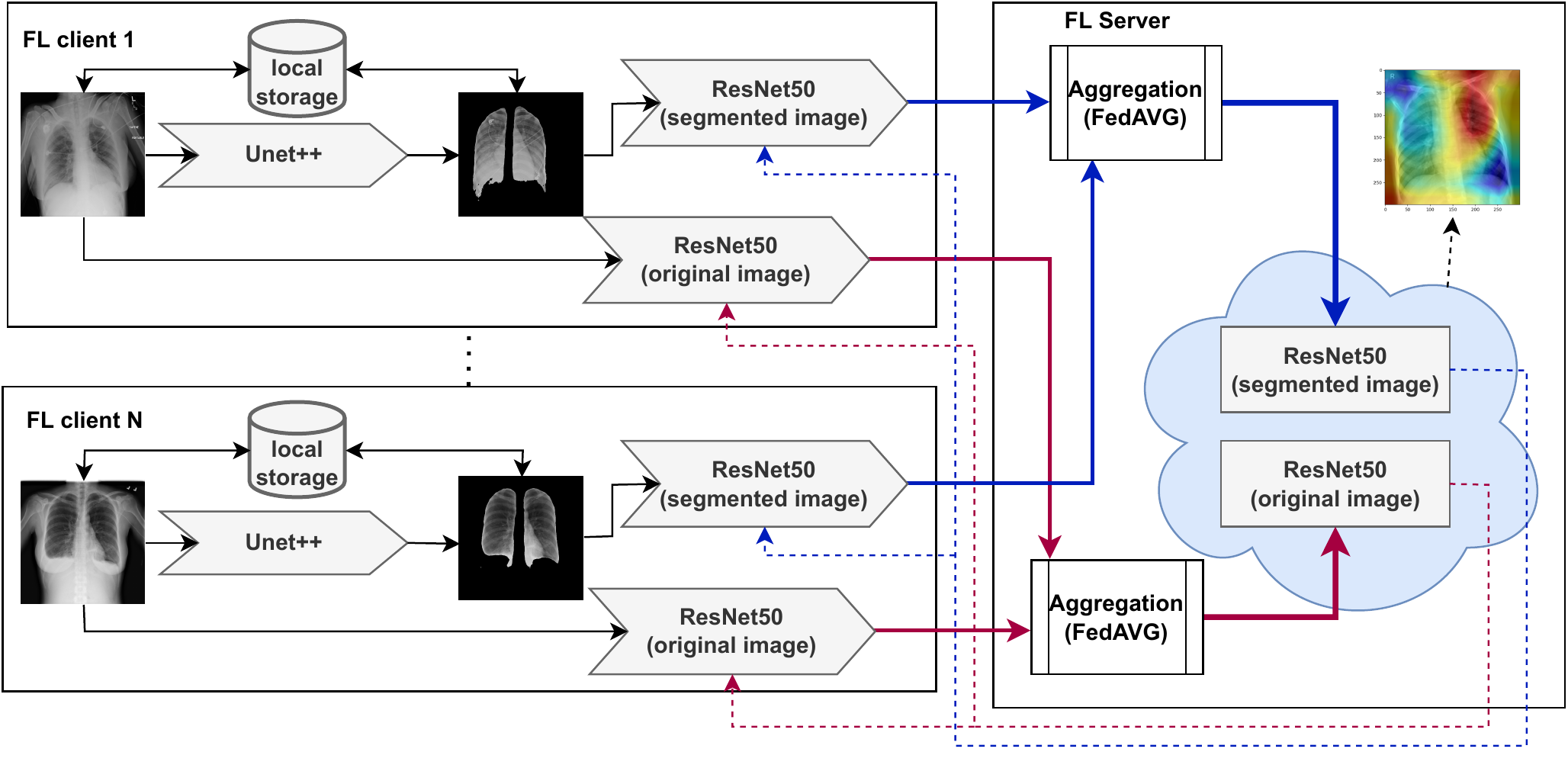}
    \caption{Methodology: combining segmentation and classification in FL setting}
    \label{fig:fl_diagram}%
\end{figure}

\section{Method: FL for Segmentation and Classification}

Our method consists of federated training of segmentation and classification models. First, we train segmentation models in a federated manner. For this purpose, we utilize the UNet++ model (with an EfficientNet-B4 backbone) that is later used to prepare the input for classification models.
At the classification stage, we use the best segmentation model in terms of the chosen performance metric, and preprocess CXR images (from the training and testing set) to extract lung regions and reduce the impact of the background noise on the prediction. 
We subsequently train one model on full images and the second on segmented ones independently, all in a federated fashion. 
During each round of federated training, clients download the global model and fine-tune it with the use of locally stored data. Once all models are fine-tuned in the given round, the server aggregates weights and the next round begins. 
After the training phase, both types of models pass through the visual explanation step using GradCAM, as in \cite{Teixeira}. We test two architectures: ResNet50 and DenseNet121, both commonly used in medical image data classification~\cite{abnormality_classification}. An overview of the proposed method for classification stage is depicted in Fig~\ref{fig:fl_diagram}.

\section{Experiments and Results}

\subsection{Datasets}

\textbf{Chest X-Ray Dataset}: To train the UNet++ model in a federated manner, we use this data set, which is a union of two other data sets known as Chest X-Ray Images (Pneumonia) \cite{cxrdataset}, \cite{C19dataset}. The dataset consists of 6380 CXR images. 
\noindent
\textbf{RSNA 2018 Dataset}: To evaluate our method, we use an open-source RSNA Pneumonia Detection Challenge 2018 chest X-ray data \cite{RSNA}. In total, the dataset consists of 26684 CXR images in the DICOM format. There are 3 classes in the dataset: "Normal" (8525 - train/326 - test), "No Lung Opacity / Not Normal" (11500 - train/321 - test) and "Lung Opacity" (5659 - train/353 - test).

\subsection{Implementation details}

We implement our models in Python 3.8 with the PyTorch v.1.10.1 and Flower v.0.17.0  frameworks, based on our previous experience~\cite{flower}. We train our models on 4 nodes of a cluster with 1 $\times$ NVIDIA v100 GPU each.

For the \textbf{segmentation task}, we use UNet++ with EfficientNet-B4 backbone pretrained on ImageNet. Adagrad is utilised as an optimizer for clients. We use a batch size of 2 and set learning rate and weight decay to $lr = 1 \times 10^{-3}$, $wd = 0$ respectively. We assess Jaccard score and BCE-Dice loss on a test set on the central server. The data set used to train the segmentation model was split into a training set and a test set with a 9:1 ratio, maintaining IID distribution of samples. Images are rescaled to $1024 \times 1024 $ px and augmented with random flip and random affine transformations. The central model is evaluated on a server-side test set after each training round. 
For the \textbf{classification task}, we use Adam optimizer with learning rate $lr = 1 \times 10^{-4}$ and weight decay $wd = 1 \times 10^{-5}$, and set batch size to 8. Images are rescaled to $224 \times 224 $ px and augmented with random flip and random affine transformations. We evaluate accuracy and CE loss on the test sets (segmented/non-segmented) on the central server. In both tasks, the models are pretrained on the ImageNet dataset. Such pretrained models are downloaded by clients during the first round of the process. We use the FedAvg~\cite{fedavg} aggregation strategy and split data in the IID manner among FL clients both in segmentation and classification.

\subsection{Segmentation results}

\begin{figure*}[t!]
  \centering
\begin{tabular}{cc}
\includegraphics[width=0.43\textwidth]{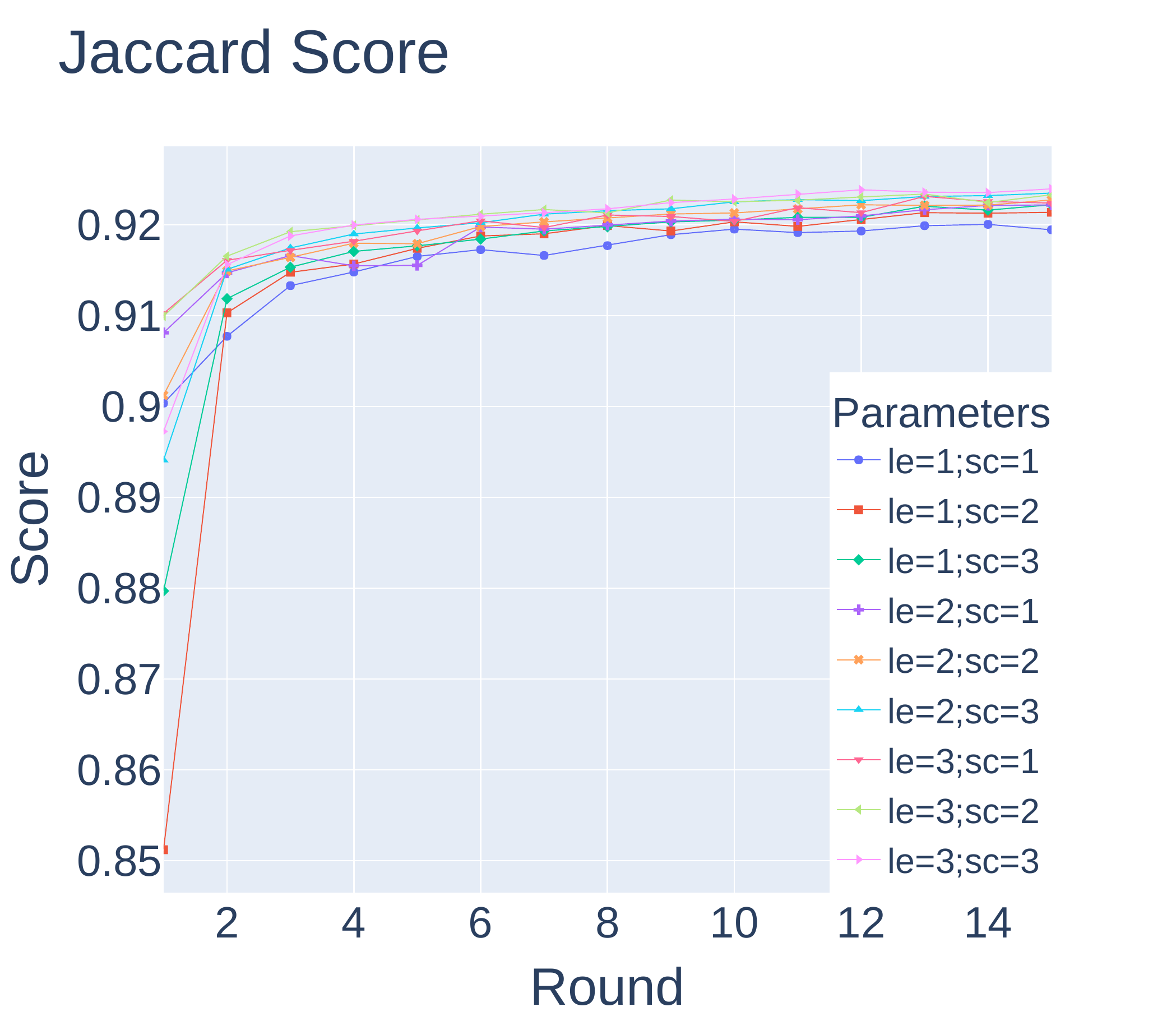}&
\includegraphics[width=0.43\textwidth]{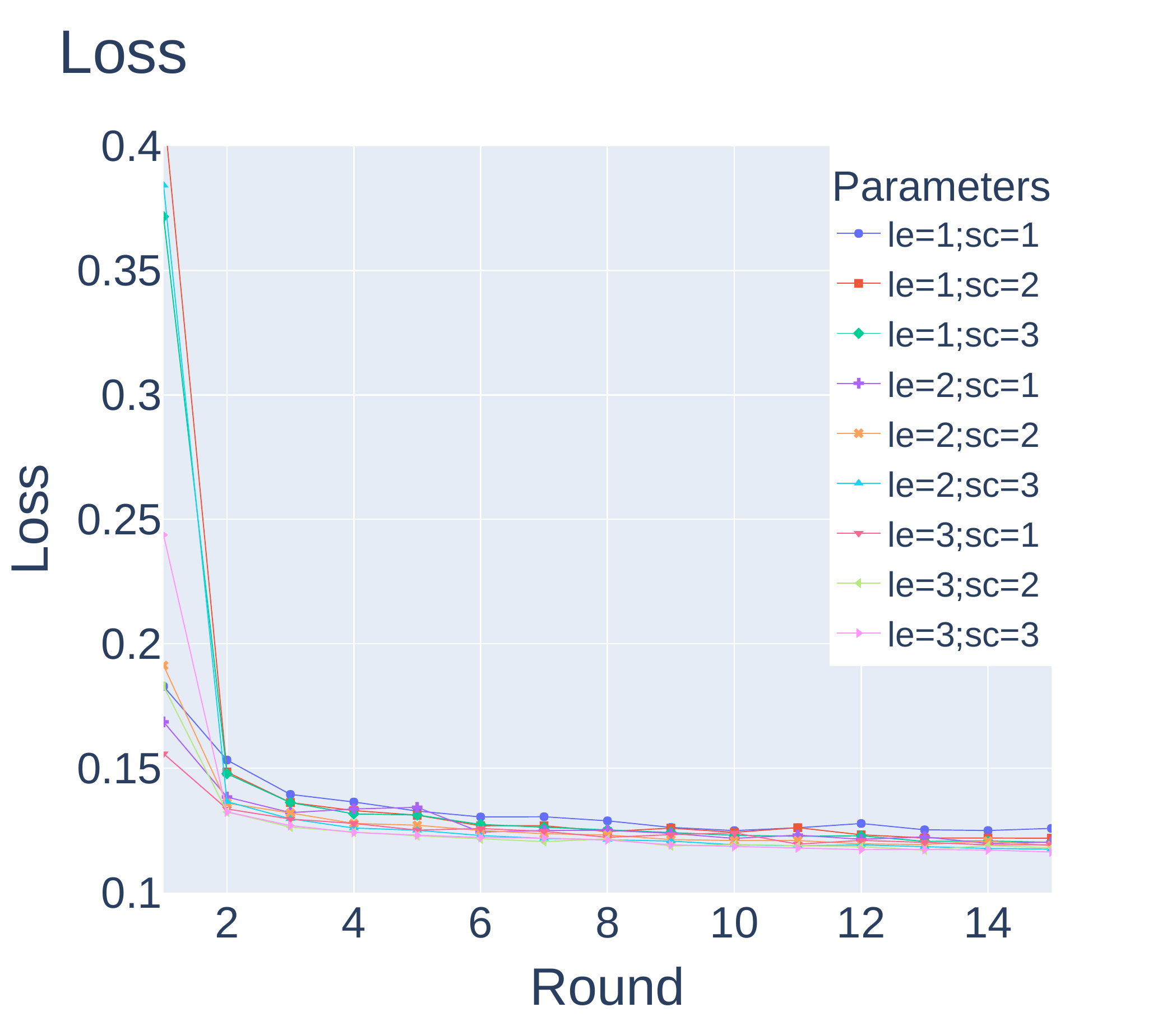} \\
a) & b) \\
\end{tabular}
    \caption{(a) Jaccard score for the test dataset, achieved by segmentation models, and (b) loss of segmentation models for the test dataset, in successive rounds of training. ”sc” - the number of clients selected by the server in each round, ”le” - the number of local epochs performed by each client per round.}
    \label{fig:perf_seg}
\end{figure*}

In order to find the optimal central segmentation model, we evaluate several configurations of parameters typical for FL such as the number of local epochs performed by each client during every training round and the fraction of clients selected by the server during each round. The process of training each model consists of 15 rounds. The Jaccard score and loss obtained by each model are presented in Fig \ref{fig:perf_seg}. For each configuration, we check the number of rounds required to achieve a Jaccard score of 0.92 twice. Results are presented in Table \ref{tab:jaccard_score_rounds_table}. We identify that for a fixed number of local epochs, a greater fraction of selected clients results in a smaller number of rounds needed to exceed the score of 0.92, similarly to the trend observed in \cite{fedavg}. The highest score (0.924) is achieved by the model trained with 3 local epochs and 3 selected clients in the 15th round of training. This model is later used to generate masks for classification. 

\begin{table}[ht!]
\centering
\caption{Number of rounds needed by the segmentation model to exceed a Jaccard Score of 0.92 for the serverside test dataset. "sc" - the number of the clients selected by the server in each round, "le" - the number of local epochs performed by each client per round.}
\begin{scriptsize}
\label{tab:jaccard_score_rounds_table}
\begin{tabular}{|c|c|c|}
\hline
\textbf{Configuration} & \textbf{Experiment 1} & \textbf{Experiment 2} \\ \hline\hline
le = 1 \& sc = 1                             & 13                                         & 14       \\ \hline
le = 1 \& sc = 2                             & 11 & 10             \\ \hline
le = 1 \& sc = 3                             & \textbf{9}   & \textbf{9}             \\ \hline\hline
le = 2 \& sc = 1                             & 9   & 9              \\ \hline
le = 2 \& sc = 2                             & 7   & 7         \\ \hline
le = 2 \& sc = 3                             & \textbf{6}   & \textbf{6}        \\ \hline\hline
le = 3 \& sc = 1                             & 6                                          & 6              \\ \hline
le = 3 \& sc = 2                             & 5                                          & 5      \\ \hline
le = 3 \& sc = 3                             & \textbf{5}  & \textbf{5}               \\ \hline
\end{tabular}
\end{scriptsize}
\end{table}

\subsection{Classification results}

\begin{figure*}[t!]
  \centering
\begin{tabular}{cc}
\includegraphics[width=0.5\textwidth]{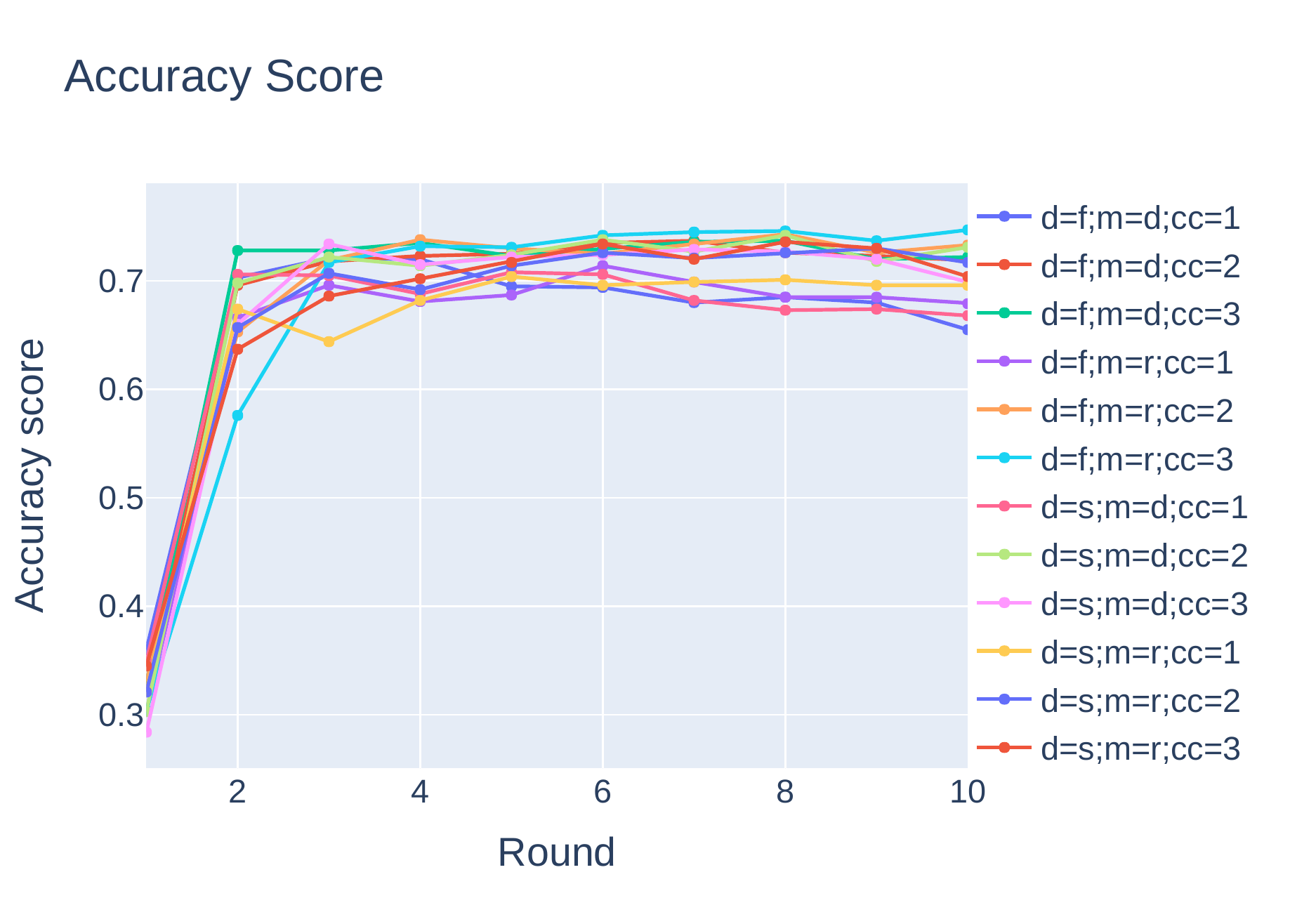}&
\includegraphics[width=0.5\textwidth]{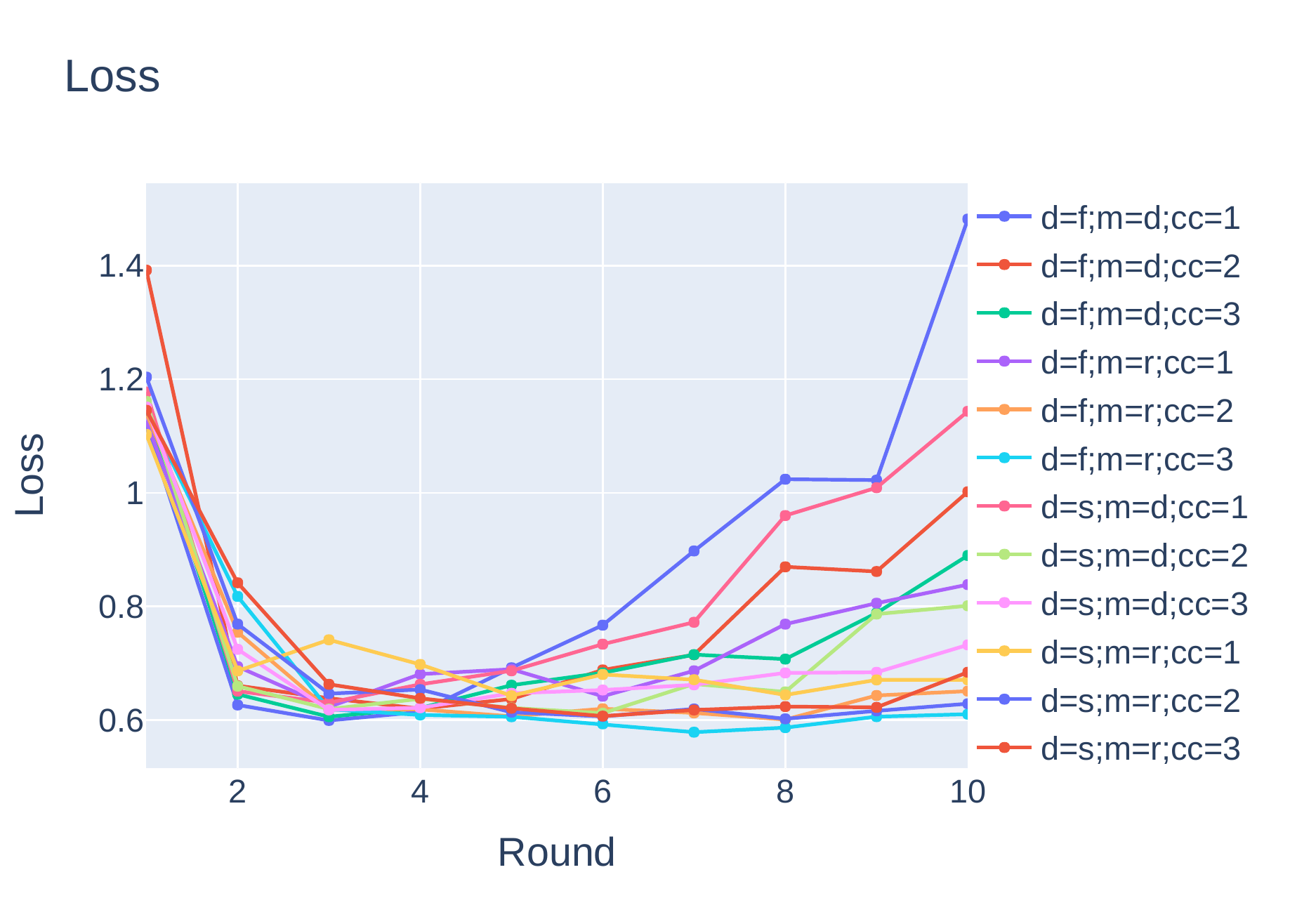} \\
a) & b) \\
\end{tabular}
    \caption{(a) Accuracy score achieved for the test dataset by classification models, and (b) loss of classification models on test dataset, in successive rounds of training. "d" defines dataset type (f - full/s - segmented), "m" defines model (d - DenseNet121, r - ResNet50), "cc" indicates number of clients participating in training.}
    \label{fig:perf_class}
\end{figure*}

In the case of the classification task, we evaluate how splitting the same amount of training data between 1, 2 and 3 clients impacts global model quality. Additionally, we assess differences between results obtained with ResNet50 and DenseNet121 architectures on full and segmented images. The accuracy score and loss for 10 rounds of training are presented in Fig. \ref{fig:perf_class}. It can be noted that the training process overfits in the case of 1 client and DenseNet121 model, both for segmented and full images, which is represented by a high loss value in the two last rounds for those configurations. The degradation of the global model quality can be also observed for DenseNet121 trained with full images on 2 and 3 clients. The lowest and most stable loss values are obtained for the ResNet50 model trained with 2 and 3 clients for full images and 1 to 3 clients for segmented images. Table \ref{tab:classification_scores} presents maximum accuracy and minimum loss values for each configuration of model architecture and dataset type. 

\begin{figure}[t!]
\centering
\begin{tabular}{cccc}

\includegraphics[width=2.4cm]{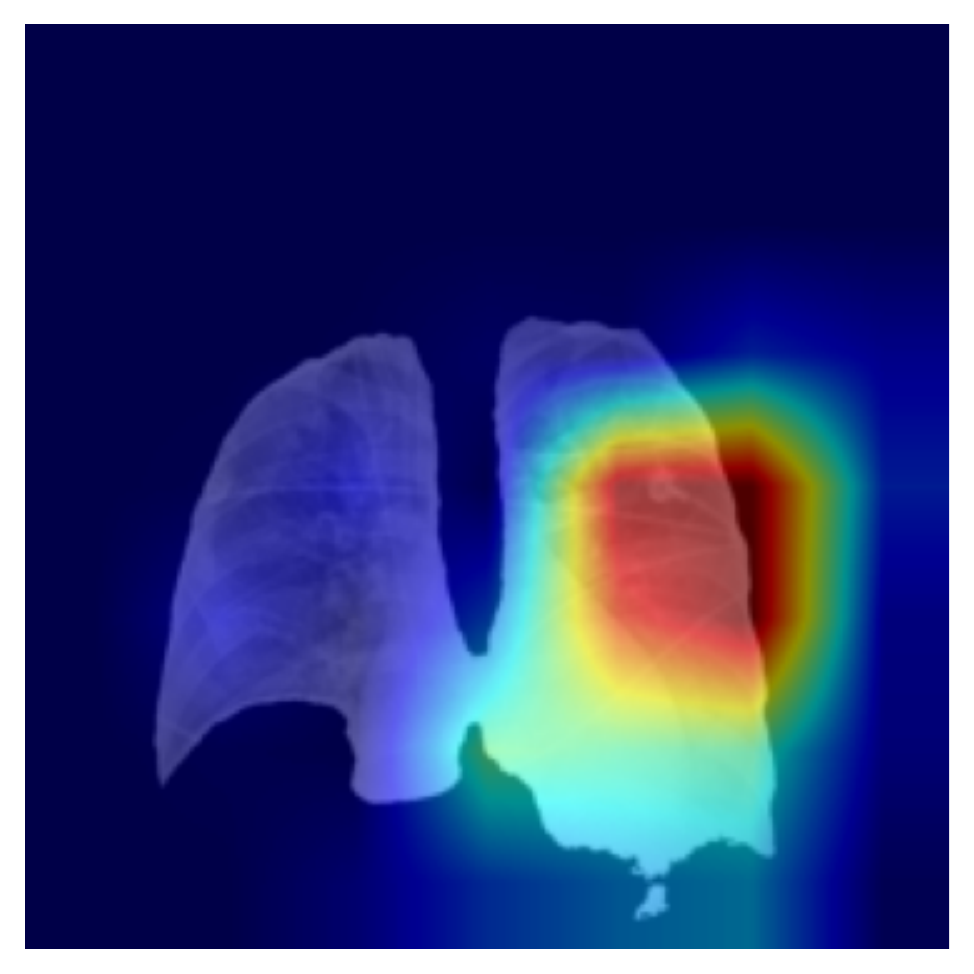}&
\includegraphics[width=2.4cm]{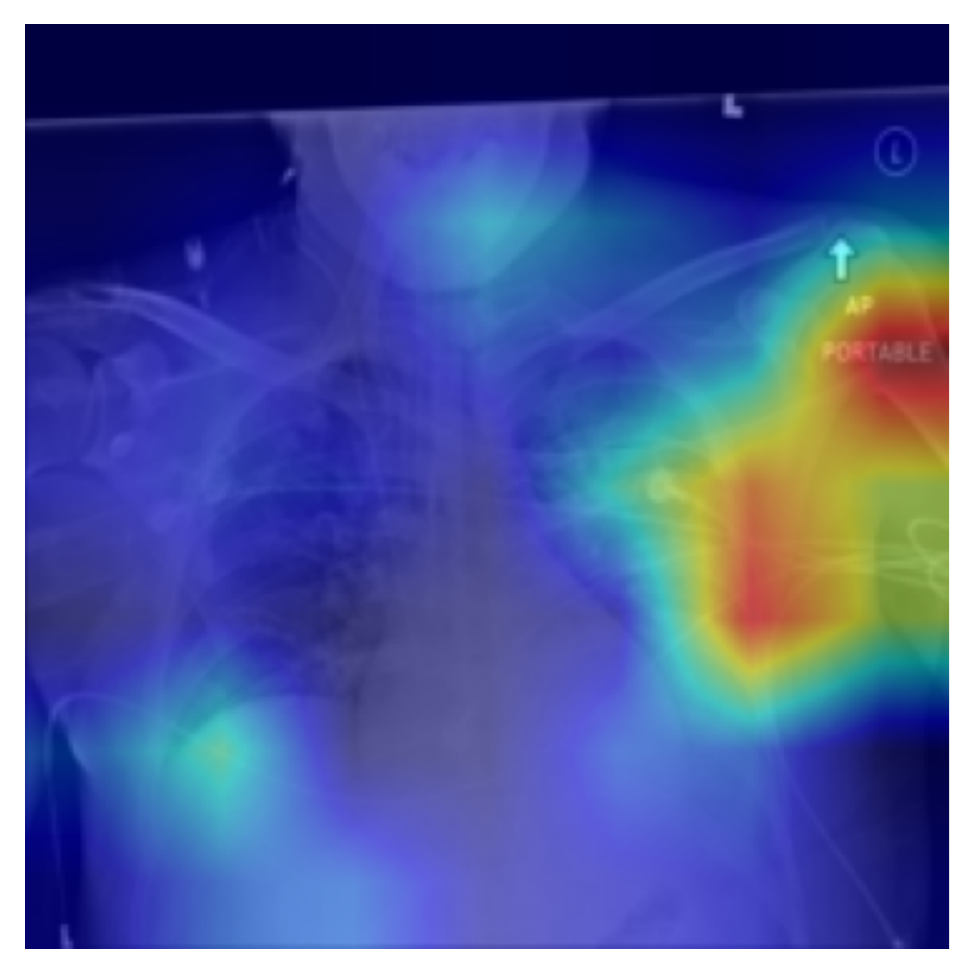}&
\includegraphics[width=2.4cm]{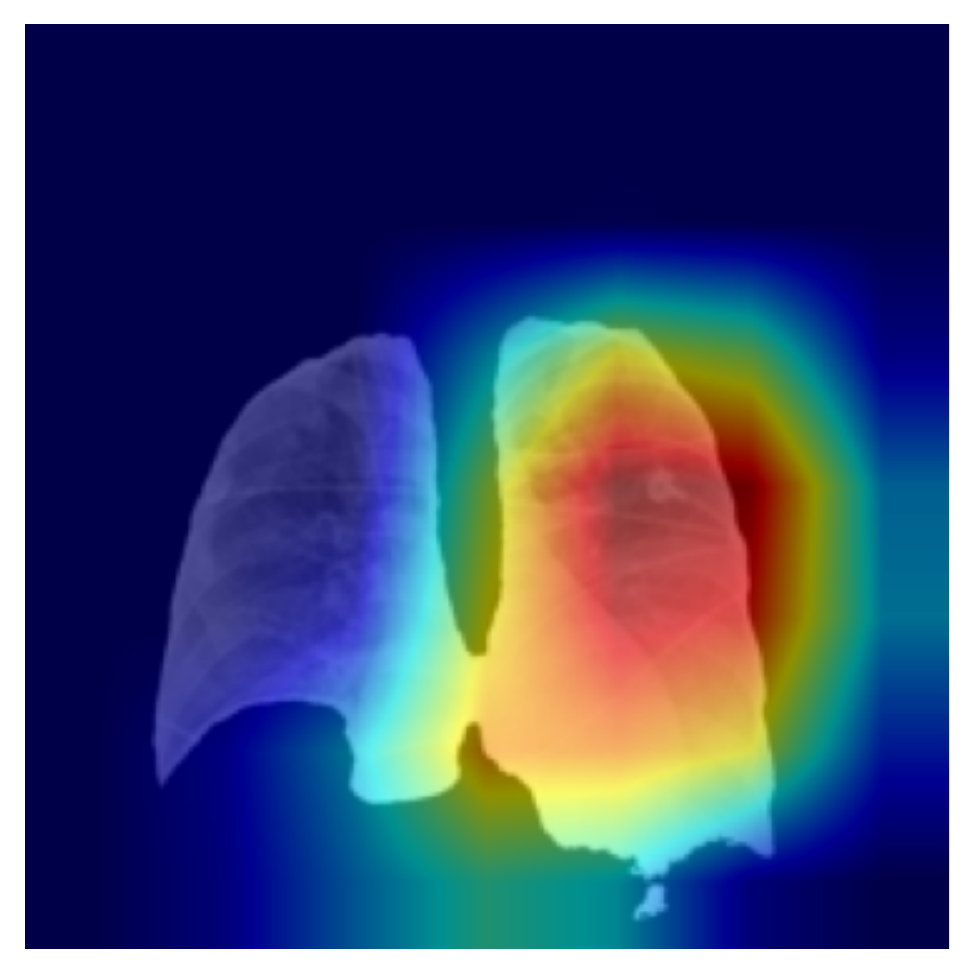}&
\includegraphics[width=2.4cm]{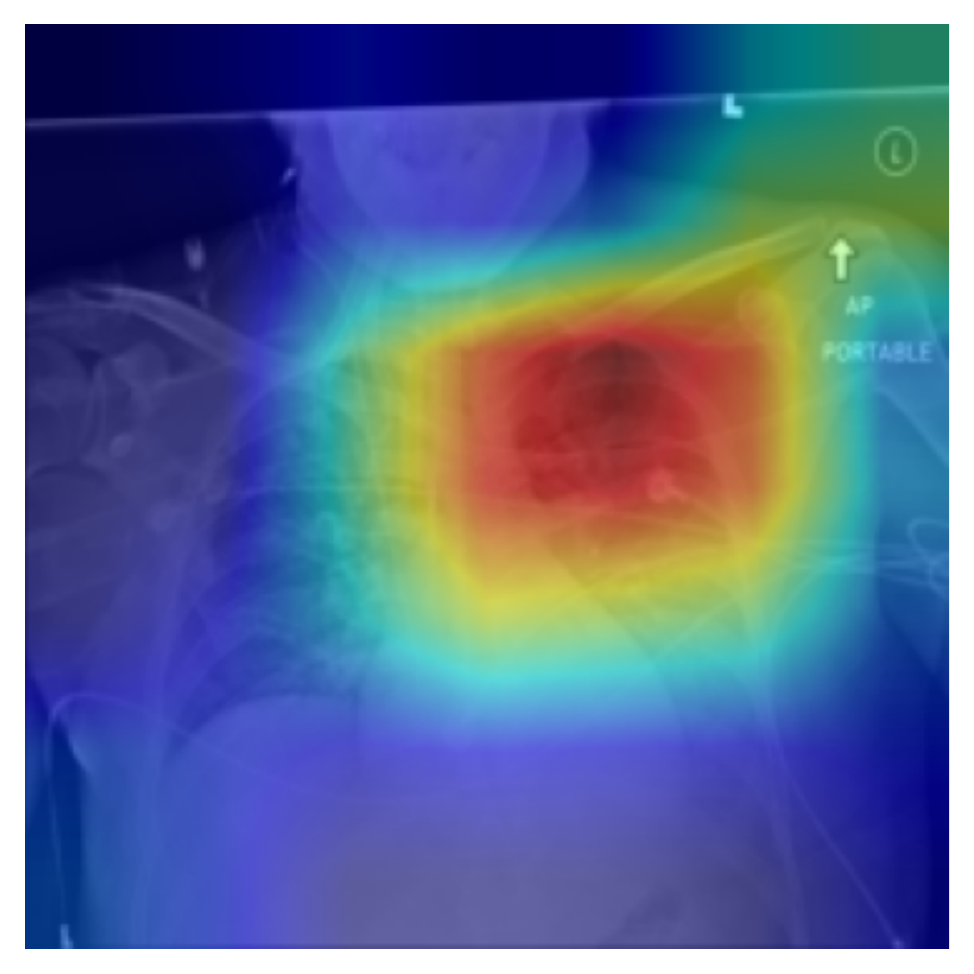} \\

\includegraphics[width=2.4cm]{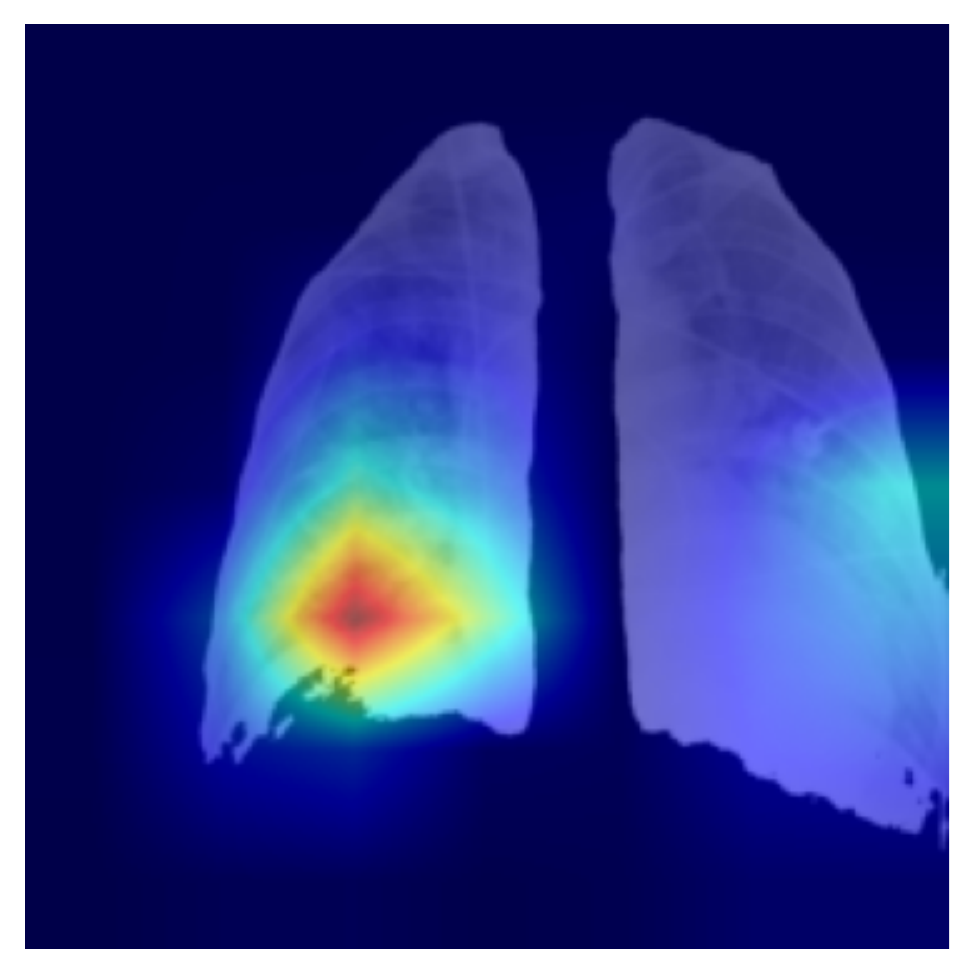}&
\includegraphics[width=2.4cm]{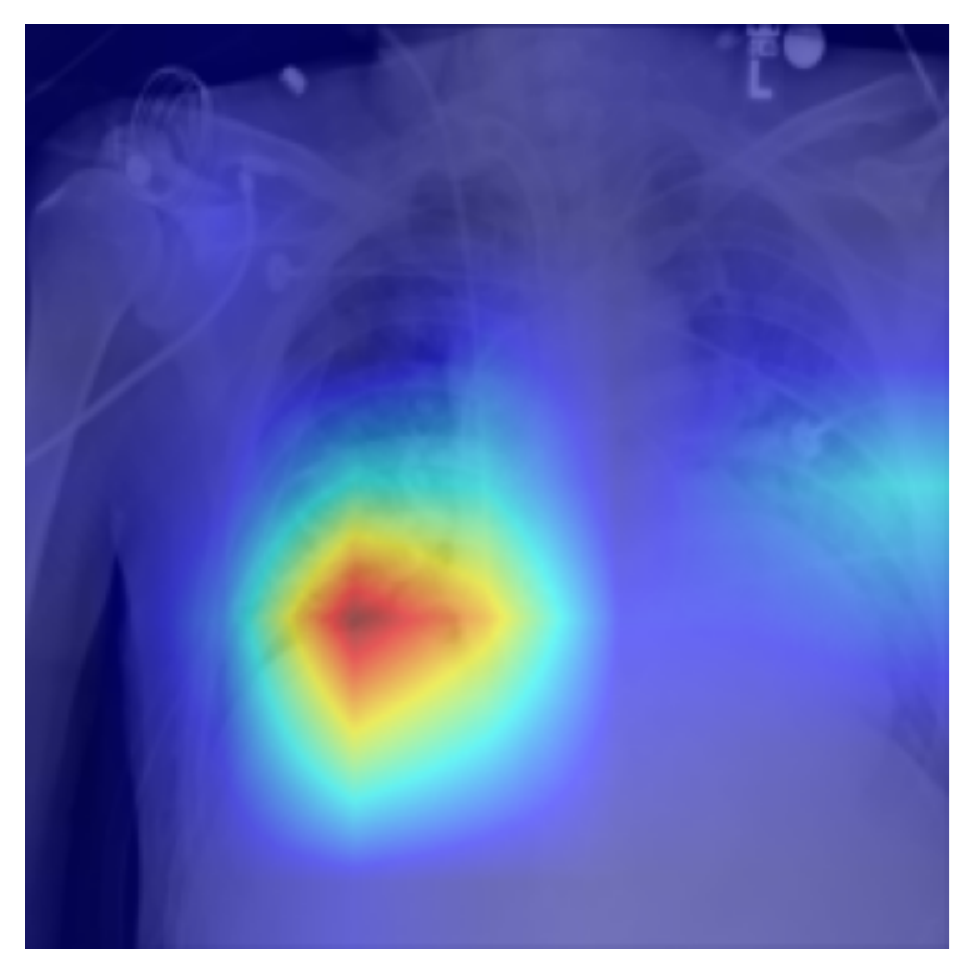}&
\includegraphics[width=2.4cm]{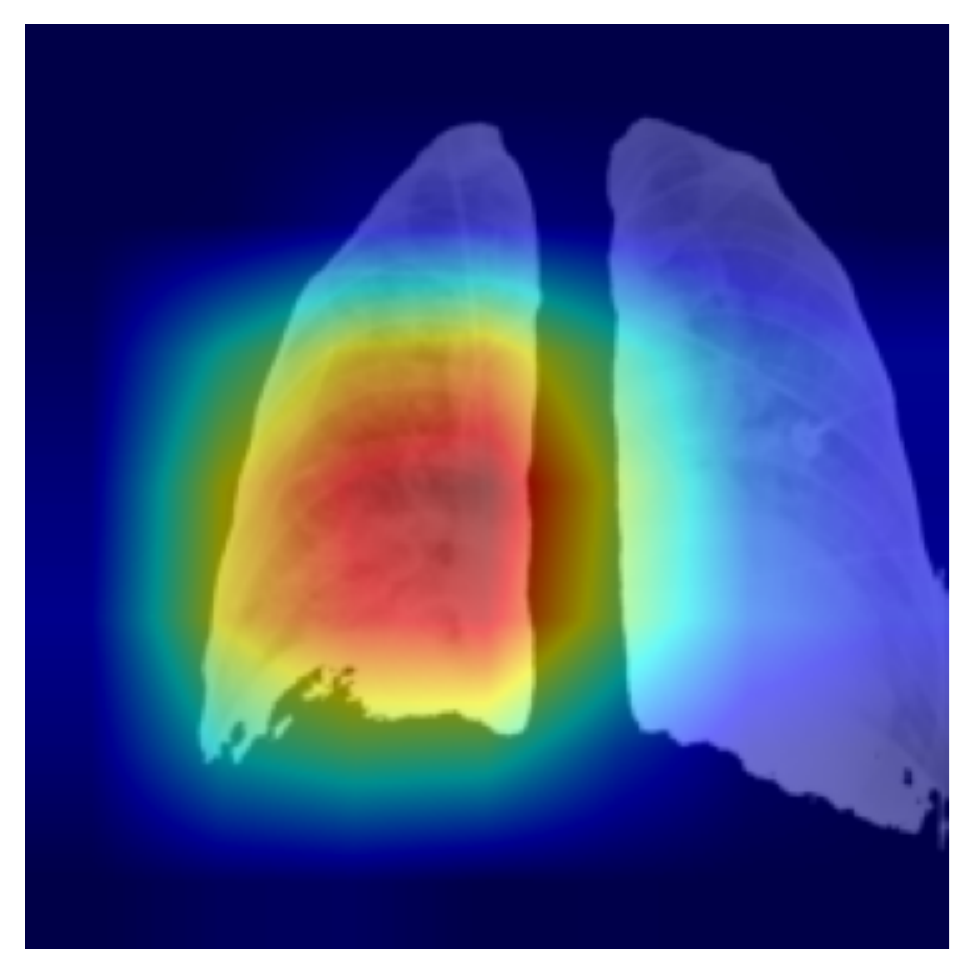}&
\includegraphics[width=2.4cm]{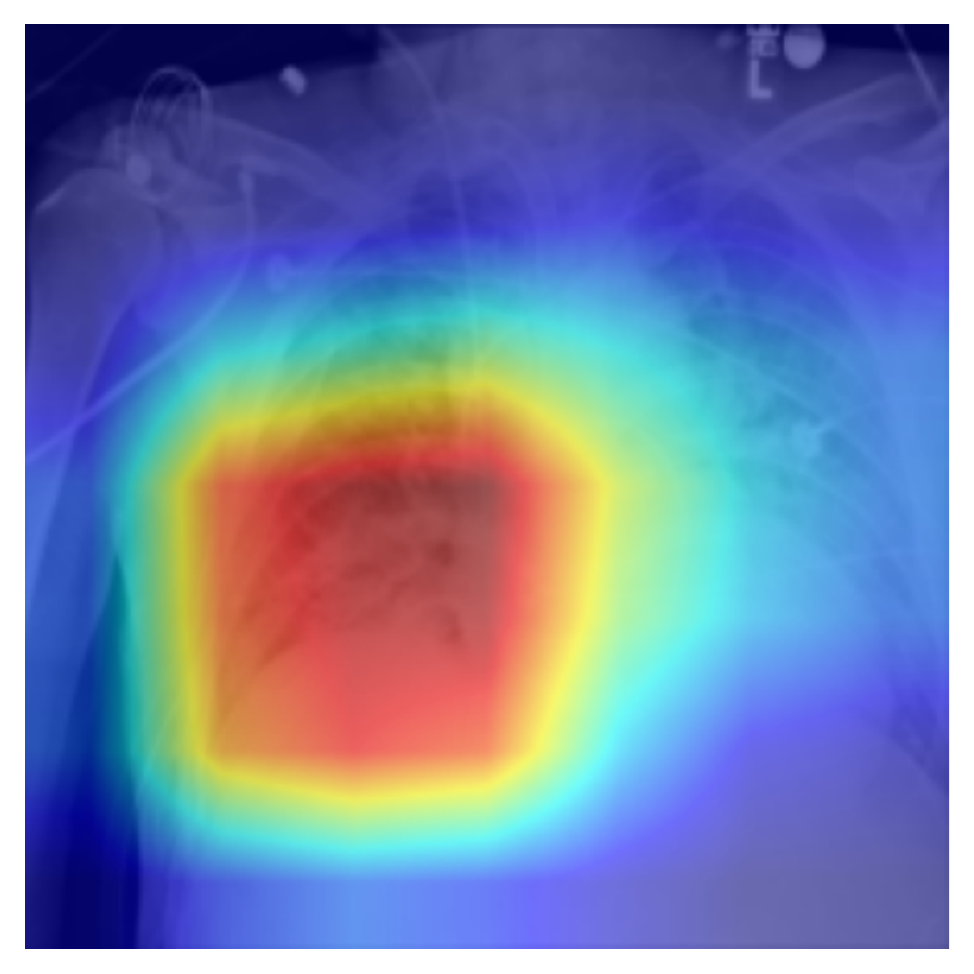} \\

a) Resnet50 &  b) Resnet50 & c) DenseNet121 & d) DenseNet121 \\ \hspace{2.2mm} (segmented) & &  \hspace{1.5mm}(segmented) &  \\
\end{tabular}
    \caption{Grad-CAM visualisations of Lung Opacity samples. In some instances, segmentation resulted in activations focused more on the lung area (upper sample). However, for a majority of cases, visualisation was comparable for segmented and full images (lower sample).}
    \label{fig:gradcam}
\end{figure}


The best accuracy, $0.757$, is achieved for ResNet50 model trained on two clients. The worst-performing model is DenseNet121 trained on full images on a single client. In general, the evaluation shows that training on a single client results in overall worse accuracy compared to training with 2 and 3 clients, which is reflected in Fig~\ref{tab:classification_scores}. This leads to the conclusion that in this case, splitting the data among distinct clients and training the model in the FL manner helps maintain generalizability and avoid overfitting. 
We observe that models trained on segmented images perform consistently worse than models trained on full images, as is the case for~\cite{Teixeira}. There is one exception: for the DenseNet121 model the best accuracy is achieved for segmented images (0.742).

To understand qualitative differences in the classification of segmented and full images, we perform Grad-CAM visualisation for ResNet50 and DenseNet121 models. We identify samples that show that the use of segmented images leads to activations more focused on the lung area (as presented in the upper sample in Fig. \ref{fig:gradcam}), which is beneficial for model interpretability. However, it can be observed that samples in which the activations are already focused on regions with pathological lung changes, for both full and segmented images, are prevalent. We believe that the small difference in the quality of the models trained on full and segmented images can be explained by the common presence of that similarity.

\begin{table}[ht!]
\centering
\caption{Maximum accuracy and minimum loss values obtained for each classification model on the test set. "d" defines dataset type (f - full/s - segmented), "m" defines model (d - DenseNet121, r - ResNet50), "cc" indicates number of clients participating in training. Values listed in boldface correspond to extremes in each \textit{model} / \textit{dataset kind} subset.}
\label{tab:classification_scores}
\begin{scriptsize}
\begin{tabular}{|c|c|c|}
\hline
\textbf{Configuration} & \textbf{Max. accuracy} & \textbf{Min. loss} \\ \hline\hline
d=f \& m=d \& cc=1     & 0.721                  & \textbf{0.599}     \\ \hline
d=f \& m=d \& cc=2     & \textbf{0.737}         & 0.620              \\ \hline
d=f \& m=d \& cc=3     & \textbf{0.737}         & 0.606              \\ \hline\hline
d=f \& m=r \& cc=1     & 0.714                  & 0.623              \\ \hline
d=f \& m=r \& cc=2     & \textbf{0.757}         & 0.601              \\ \hline
d=f \& m=r \& cc=3     & 0.747                  & \textbf{0.579}     \\ \hline\hline
d=s \& m=d \& cc=1     & 0.708                  & 0.631              \\ \hline
d=s \& m=d \& cc=2     & \textbf{0.742}         & \textbf{0.612}     \\ \hline
d=s \& m=d \& cc=3     & 0.734                  & 0.618              \\ \hline\hline
d=s \& m=r \& cc=1     & 0.704                  & 0.643              \\ \hline
d=s \& m=r \& cc=2     & 0.730                  & \textbf{0.602}     \\ \hline
d=s \& m=r \& cc=3     & \textbf{0.736}         & 0.607              \\ \hline
\end{tabular}
\end{scriptsize}

\end{table}

\section{Conclusions}

In this paper, we evaluated deep learning-based models in the context of CXR image analysis. We conducted experiments in a FL environment to understand the impact of FL-related parameters on the global model performance in segmentation and classification tasks. We also prepared Grad-CAM visualisations for classification models.
We found that in the segmentation task, when the number of local epochs is fixed, the model reaches the desired quality faster with a greater fraction of selected clients. In addition, setting a greater number of local epochs for each client also leads to the same behaviour, which may contribute to lower network traffic in FL processes. Moreover, we conclude that splitting the same dataset among distinct FL clients may lead to improvements in classification for the tested models. We observed a higher accuracy score for full images compared to segmented images in the classification task. However, models trained on segmented images may be characterized by improved interpretability. 



\section*{Acknowledgements}

This publication is partly supported by the EU H2020 grant Sano (No. 857533) and the IRAP Plus programme of the Foundation for Polish Science. This research was supported in part by the PL-Grid Infrastructure. We would like to thank Piotr Nowakowski for his assistance with proofreading the manuscript.

%
%
%

\begin{thebibliography}{8}



\bibitem{Teixeira} Teixeira, L.O., et al.
: Impact of Lung Segmentation on the Diagnosis and Explanation of COVID-19 in Chest X-ray Images. Sensors. 21, 7116 (2021). https://doi.org/10.3390/s21217116.

\bibitem{Sheller} Sheller, M.J., et al.:
Federated learning in medicine: facilitating multi-institutional collaborations without sharing patient data. Sci Rep. 10, 12598 (2020). https://doi.org/10.1038/s41598-020-69250-1.

\bibitem{Chen} Chen, Z., et al.: Personalized Retrogress-Resilient Framework for Real-World Medical Federated Learning. In: 
Medical Image Computing and Computer Assisted Intervention – MICCAI 2021. pp. 347–356. Springer 
(2021). https://doi.org/10.1007/978-3-030-87199-4\_33.

\bibitem{Wu} Wu, Y., et al.: Federated Contrastive Learning for Volumetric Medical Image Segmentation. In: 
Medical Image Computing and Computer Assisted Intervention – MICCAI 2021. pp. 367–377. Springer 
(2021). https://doi.org/10.1007/978-3-030-87199-4\_35.

\bibitem{Dong} Dong, N., Voiculescu, I.: Federated Contrastive Learning for Decentralized Unlabeled Medical Images. In: 
Medical Image Computing and Computer Assisted Intervention – MICCAI 2021. pp. 378–387. Springer 
(2021). https://doi.org/10.1007/978-3-030-87199-4\_36.

\bibitem{flower} Jabłecki, P., et al.: Federated Learning in the Cloud for Analysis of Medical Images - Experience with Open Source Frameworks. In: 
Clinical Image-Based Procedures, Distributed and Collaborative Learning, Artificial Intelligence for Combating COVID-19 and Secure and Privacy-Preserving Machine Learning. pp. 111–119. Springer
(2021). https://doi.org/10.1007/978-3-030-90874-4\_11.


\bibitem{RSNA} 
Shih, G.
et. al.
: Augmenting the National Institutes of Health Chest Radiograph Dataset with Expert Annotations of Possible Pneumonia. Radiology: Artificial Intelligence. 1, e180041 (2019). https://doi.org/10.1148/ryai.2019180041.

\bibitem{cxrdataset} Kermany, D., et al.:
Labeled Optical Coherence Tomography (OCT) and Chest X-Ray Images for Classification. 2, (2018). https://doi.org/10.17632/rscbjbr9sj.2.


\bibitem{C19dataset} Cohen, J.P., et al.: COVID-19 Image Data Collection: Prospective Predictions Are the Future. arXiv:2006.11988 [cs, eess, q-bio]. (2020).

\bibitem{abnormality_classification} Tang, Y.-X., et al.: Automated abnormality classification of chest radiographs using deep convolutional neural networks. npj Digit. Med. 3, 70 (2020). https://doi.org/10.1038/s41746-020-0273-z.

\bibitem{fedavg} McMahan, H.B., et el.: Communication-Efficient Learning of Deep Networks from Decentralized Data. (2016). https://doi.org/10.48550/ARXIV.1602.05629.

\bibitem{kaissis2020secure} Kaissis, G. A. et al: Secure, privacy-preserving and federated machine learning in medical imaging. 
Nature Machine Intelligence 2 (6) (2020) 305–311. 25

\end{thebibliography}
%

\end{document}